\documentclass[twocolumn,prl,showpacs,amsmath,amssymb,superscriptaddress]{revtex4}
\usepackage{graphicx,color}
\usepackage{hyperref}
\begin{document}
\title{Fractional Shot Noise in the Kondo Regime}

\author{Eran Sela}
\affiliation{Department of Condensed Matter Physics, Weizmann
Institute of Science, Rehovot, 76100, Israel}

\author{Yuval Oreg}
\affiliation{Department of Condensed Matter Physics, Weizmann
Institute of Science, Rehovot, 76100, Israel}

\author{Felix von Oppen}
\affiliation{Institut f\"ur Theoretische Physik, Freie Universit\"at
Berlin, Arnimallee 14, 14195 Berlin, Germany}

\author{Jens Koch}
\affiliation{Institut f\"ur Theoretische Physik, Freie Universit\"at
Berlin, Arnimallee 14, 14195 Berlin, Germany}

\date{\today}

\begin{abstract}
Low temperature transport through a quantum dot in the Kondo
regime proceeds by a universal combination of elastic and
inelastic processes, as dictated by the low-energy Fermi-liquid
fixed point. We show that as a result of inelastic processes, the
charge detected by a shot-noise experiment is enhanced relative to
the noninteracting situation to a universal fractional value,
$e^*=5/3 e$. Thus, shot noise reveals that the Kondo effect
involves many-body features even at low energies, despite its
Fermi-liquid nature. We discuss
the influence of symmetry breaking perturbations.
\end{abstract}

\maketitle {\it Introduction}.---Shot-noise measurements in
mesoscopic devices provide a direct measurement of the effective
charge $e^*$ of the current-carrying particles. Prominent examples
in which this charge differs from the electron charge $e$ include
the observation of the fractional charge $e^*=e/3$ in the
fractional quantum Hall regime~\cite{PicciottoSaminadayar97}, as
well as the detection of the Cooper-pair charge $e^*=2e$ in normal
metal-superconductor junctions~\cite{Lefloch03}. In this paper we
study shot noise in quantum dots in the Kondo limit. Despite the
Fermi liquid (FL) nature of the low-energy fixed point of the
Kondo effect, we find that the effective backscattering charge is
a universal quantity satisfying $e^* >e$. Unlike in quantum Hall
systems and superconductors, this enhancement relative to the
noninteracting value is {\it not} related to the fundamental
quasiparticle charge. Instead, it is a direct consequence of
interactions between quasiparticles of charge $e$, which lead to
simultaneous backscattering of two quasiparticles.

The Kondo effect occurs in quantum dots \cite{Glazman05} when the
dot carries an effective spin, and charge fluctuations are frozen
out by the strong Coulomb repulsion. Virtual tunneling of
electrons into and out of the dot induces an antiferromagnetic
coupling of the dot spin with the electrons in the leads. In this
paper, we focus on the regime of temperatures $T$ well below the
Kondo temperature $T_K$, where the dot spin is locked into a
singlet state with the lead electrons. Then, for two leads coupled
symmetrically to the dot, the linear-response conductance is
enhanced to the maximal unitary value $g_0 = 2
e^2/h$~\cite{Wielshort00}, corresponding to the conductance of a
fully transparent channel with transmission probability
$T(\epsilon) = 1$.

Shot noise in the Kondo effect was recently addressed
theoretically for a wide range of temperatures and voltages
V~\cite{Schiller98,Meir02}. For energies well above $T_K $, shot
noise exhibits the typical enhancement $\propto
\log^{-2}(eV/T_K)$, then it develops a peak around~$T_K$, and is
finally suppressed at low energies. The low-temperature
suppression can be understood from the expression for the shot
noise $S$ in noninteracting systems~\cite{Lesovik89}
\begin{equation}
\label{eq:non} S= 2 g_0 \int d\epsilon T(\epsilon) [1-T(\epsilon)]
[f_s(\epsilon)-f_d(\epsilon)],
\end{equation}
which vanishes in the unitary limit $T(\epsilon) \rightarrow 1$.
Here $f_s$ and $f_d$ denote the Fermi distribution functions of the
source and drain, respectively. Intuitively, while the incident
fermionic carrier flow is fluctuationless at zero temperature, the
transmitted and reflected carrier flows generally exhibit
probabilistically generated noise. However, if $T(\epsilon) = 1$ or
$0$, no noise is generated by the scatterer.

The starting point of the present paper is the observation that
close to the limit of perfect transmission, it is natural to
extract the charge of the {\it backscattered} particles from the
ratio
\begin{equation}
\label{eq:qef} e^*=
 {S}/{2 I_b},
\end{equation}
where $I_b$ denotes the backscattering current of reflected
carriers. Indeed, this definition was used to extract the
quasiparticle charge in the fractional quantum Hall
effect~\cite{PicciottoSaminadayar97}. In the noninteracting case,
$I_b = 2\frac{e}{h} \int d\epsilon [1-T(\epsilon)]
[f_s(\epsilon)-f_d(\epsilon)]$, and using Eqs.~(\ref{eq:non}) and
(\ref{eq:qef}) we have $e^* = e$ when $T (\epsilon )\rightarrow 1$
for energies $\epsilon$ close to the Fermi energy. In contrast, it
is the central result of this paper that $e^* =\frac{5}{3} e$ in
the Kondo regime. We show below that this is a universal property
of the Kondo effect, which is independent of the Kondo temperature
$T_K$.

\begin{figure}[t]
\begin{center}
\includegraphics*[width=90mm,height=60mm]{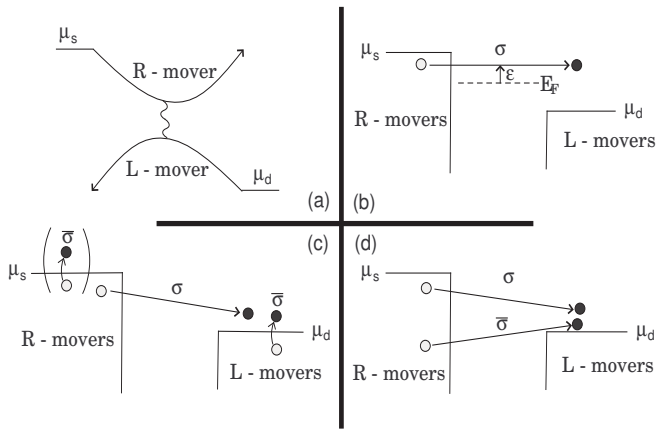}
\caption{Transport processes near the unitary limit. (a) In the
unitary limit an electron incident from the source is almost
totally transmitted into the drain, a process describing the free
motion of a R-mover. Weak backscattering between R-movers and
L-movers is denoted by the wavy line.  (b) Elastic backscattering
of one R-mover with energy $\epsilon$ relative to $E_F = (\mu_s+
\mu_d) /2$. The energy dependence of this amplitude $\propto
\alpha \epsilon / T_K$ corresponds to scattering off a resonant
level of width $\propto T_K/\alpha$, centered at $E_F$. (c)
Inelastic backscattering of one R-mover accompanied by creation of
a particle hole pair within one reservoir. (d) Inelastic
backscattering of two R-movers with opposite spins $\sigma$ and
$\bar{\sigma}$. The amplitudes of processes (c) and (d) are
proportional to $\beta$, see Eq.~(\ref{eq:hfp}). \label{fg:1}
\vspace{-.5cm}}
\end{center}
\end{figure}

Near the unitary limit, it is most convenient to describe the
system in the language of right movers (R-movers) propagating from
source (with chemical potential $\mu_s$) to drain (with chemical
potential $\mu_d$) and left movers (L-movers) propagating from
drain to source. Deviations from the unitary limit will allow
R-movers to backscatter into L-movers and vice versa, as indicated
schematically by a wavy line in Fig.~\ref{fg:1}(a). We now turn to
a discussion of the various relevant backscattering processes
which are dictated by the low-energy fixed point of the Kondo
effect and summarized pictorially in Figs.~\ref{fg:1}(b), (c), and
(d).

In a naive picture, the Kondo effect is thought of as the
formation of a single-quasiparticle resonance of width~$T_K$,
centered exactly at the Fermi energy $E_F$. Then, individual
quasiparticles are backscattered as energies $\epsilon$ away from
the Fermi energy become relevant due to finite temperature or
voltage [see Fig.~\ref{fg:1}(b)]. The rate for this process grows
quadratically in $\max \{ T/T_K, V/T_K \}$. However, as $T$ or $V$
increase, an additional inelastic channel opens for scattering
between the R-movers and L-movers, in which the backscattering
event is accompanied by the simultaneous creation of a
particle-hole pair. In this case, the corresponding rates are
again quadratic in $\max \{ T/T_K, V/T_K \}$ due to phase-space
restrictions for particle-hole-pair creation. If the particle-hole
pair is created within the drain or within the source [see
Fig.~\ref{fg:1}(c)], the process effectively backscatters a single
mover. However, when particle and hole are created in drain and
source, respectively [see Fig.~\ref{fg:1}(d)], we encounter an
event in which {\it two} R-movers backscatter simultaneously
\cite{Remark06e}. These are the processes that lead to an
effective backscattering charge $e<e^*<2e$ as measured by shot
noise.

The universality of $e^*$ is a consequence of the fact that the
Kondo resonance is tied to the Fermi level. This fixes the ratio
between the amplitudes~$\alpha$ for elastic scattering and $\beta$
for the interactions which generate inelastic scattering. It is a
central result in Nozi\`{e}res' FL theory of the Kondo
effect~\cite{Nozieres74} that $\alpha=\beta$. Thus, the
fixed-point Hamiltonian Eq.~(\ref{eq:hfp}) describes the
low-energy properties by a single parameter $T_K$. It is
interesting to note that the Wilson ratio, i.e., the ratio between
the relative changes in the susceptibility and the specific heat
due to the local spin, $W = (\delta \chi / \chi)/(\delta C_v/C_v)
=1+ \beta/ \alpha=2$, is another quantity which acquires a
universal value due to the same reason. However, the universality
of $W$ is actually restricted to situations where the $g$-factors
of localized spin and conduction electrons are
equal~\cite{Finkelstein78}. We emphasize that the universality of
$e^*$ is {\it not} subject to this restriction.

The large value of the effective charge is surprising since there
are six possible processes in which one mover is backscattered [two
elastic and four inelastic processes, see Figs.~\ref{fg:1}(b) and
\ref{fg:1}(c)] compared to only a single process of two-particle
backscattering [Fig.~\ref{fg:1}(d)]. However, the phase space for
the two-particle process is significantly enhanced by the fact that
the applied voltage acts on {\it both} particles scattered from
source to drain. Indeed, we find for $eV \gg T$ that $2/3$ of the
backscattering current is carried by two-particle processes.

{\it Calculation}.---We describe a quantum dot with symmetric dot-lead couplings
near the unitary limit in the basis of L and R-movers with energy
$\xi_k=v_F k$, spin $\sigma$, as well as creation operators
$L^\dagger_{k \sigma}$ and $R^\dagger_{k \sigma}$, respectively.
The distribution of incoming R-movers (L-movers) is dictated by the
chemical potential of the source (drain). Due to the LR-symmetry,
the low-energy Hamiltonian $H[\psi]$ can be written entirely in terms of the
symmetric combination $\psi_{k\sigma}=\frac{1}{\sqrt{2}}(L_{k \sigma}+R_{k \sigma})$.

In view of the Fermi-liquid nature of the Kondo fixed point, the
low-energy physics can be completely described by the scattering
phase shift suffered by an incoming quasiparticle ($\psi_{k
\sigma}$), combined with the quasiparticle distribution $n_\sigma$
\cite{Nozieres74}. Following Nozi\`{e}res, the low-energy
expansion of this phase shift is $\delta_{\sigma} = \frac{\alpha
\epsilon}{ T_K} - \frac{\beta n_{\bar{\sigma}}}{\nu T_K}$ where
$\bar\sigma = -\sigma$ and $\nu$ is the density of states. Notice
that the phase shift of the electrons differs by $\pi/2$ from that
of the $\psi_{k \sigma}$ particles~\cite{Affleck90}. Combining
this expansion with the floating of the Kondo resonance, i.e.,
$\delta(\delta \epsilon, n=\nu \delta \epsilon)
=\delta(\epsilon=0, n=0)$, one obtains the important FL relation
$\alpha=\beta$ mentioned above.

Equivalently, the low-temperature physics can be described in terms of the
Hamiltonian \cite{Affleck90,Glazman05}
\begin{eqnarray}
\label{eq:hfp} H = \sum_{k \sigma} \xi_k \psi^\dagger_{k \sigma}
\psi_{k \sigma} -\frac{\alpha}{2 \pi \nu T_K}\sum_{k,k' \sigma}
(\xi_k+\xi_{k'}) \psi^\dagger_{k
\sigma} \psi_{k' \sigma} \nonumber \\
+\frac{\beta}{\pi \nu^2 T_K}\sum_{k_1,k_2,k_3,k_4}
\psi^\dagger_{k_1 \uparrow} \psi_{k_2 \uparrow} \psi^\dagger_{k_3
\downarrow} \psi_{k_4 \downarrow},
\end{eqnarray}
whose $t$-matrix $t_{\sigma} =\frac{1}{ 2 \pi i \nu} (1-e^{2 i
\delta_\sigma}) \simeq -\delta_{\sigma}/\pi \nu$ reproduces the
desired phase shift.

The term $\propto \alpha$ in Eq.~(\ref{eq:hfp}) which yields the
energy dependence of the phase shift, is consistent with the
picture of a resonant level of width $T_K$ centered at $E_F$.
Since the phase shift grows by $\pi$ across the resonance, this
picture implies $\alpha \sim 1$. The term $\propto \beta$
describes the quasiparticle interactions. While the corresponding
contribution to the phase shift follows from this interaction at
the Hartree level, a treatment beyond Hartree involves inelastic
processes in which quasiparticle scattering is accompanied by the
creation of particle-hole pairs.

The current $I$ transmitted from source to drain contains a
dominant (maximal) unitary contribution $ I_u = 2 \frac{e^2}{h} V$
as well as the backscattering current, $I = I_u-I_b$. The
backscattering contribution $I_b$, describing deviations from
perfect transmission, follows from the Hamiltonian
Eq.~(\ref{eq:hfp}) by evaluating the increase in the numbers of
L-movers relative to R-movers, $I_b = \frac{e}{2} \frac{d}{dt}
(N_L-N_R)$, where $N_{a=L,R} = \sum_{k \sigma} a^\dagger_{k
\sigma} a_{k \sigma}$. The zero-frequency current fluctuations
(noise power) are defined as $S = \int dt \langle \{ \delta I(0),
\delta I(t)\} \rangle$, where $\delta I = I - \langle I \rangle$.
At zero temperature, only $I_b$ contributes to shot noise.

We calculate current and noise through the biased quantum dot by
the Keldysh technique, which allows one to couple the L- and
R-movers perturbatively in $1/T_K$. To leading order, we obtain
after lengthy but straightforward calculations~\cite{Remarka}

\begin{eqnarray} \label{eq:res} I=I_u-I_b&=&
\frac{2e^2}{h}V \left[1- \frac{\alpha^2+5 \beta^2}{12} \left( \frac{V}{T_K} \right)^2 \right], \\
S &=& \frac{4e^3}{h}V  \frac{\alpha^2+9 \beta^2}{12} \left(
\frac{V}{T_K} \right)^2.
\end{eqnarray}
Substituting these expressions into Eq.~(\ref{eq:qef}), and using
$\alpha =\beta $, we indeed find $e^* =\frac{5}{3} e$.

{\it Interpretation}.---We can gain physical understanding of this
result by rewriting the Hamiltonian Eq.\
(\ref{eq:hfp}) in terms of the operators of left and right movers.
This allows us to unravel the nature of the backscattering of
R-movers into L-movers and to obtain
the rates of the various backscattering processes
from the relevant amplitudes combined with the voltage-dependent phase space.

Substituting $\psi_{k\sigma}=\frac{1}{\sqrt{2}}(L_{k \sigma}+R_{k \sigma})$,
the inelastic term $\propto \beta$ in the Hamiltonian Eq.~(\ref{eq:hfp})
takes the form $H_\beta = \frac{\beta}{4 \pi \nu^2
T_K}\sum_{k_1,k_2,k_3,k_4}^{a,b,c,d=L,R} a^\dagger_{k_1 \uparrow}
b_{k_2 \uparrow} c^\dagger_{k_3 \downarrow} d_{k_4\downarrow}$.
Clearly, it contains processes in which $0,1,$ or $2$ particles are
backscattered. An interesting process without net
backscattering is $L^\dagger_{k_1 \uparrow} R_{k_2 \uparrow}
R^\dagger_{k_3 \downarrow} L_{k_4 \downarrow}$, which contributes
to the spin current~\cite{Kindermann05} but not to the charge current
considered here.

Backscattering of {\it two} R-movers arises from the terms $ \propto
\sum L^\dagger_{k_1 \uparrow} R_{k_2 \uparrow} L^\dagger_{k_3
\downarrow} R_{k_4\downarrow}$ in $H_\beta$ [see
Fig.~\ref{fg:1}(d)]. Their contribution $I_{\beta 2} = \Gamma_{\beta
2} \cdot 2e$ to the backscattering current $I_b$ is determined by
the rate \widetext
\begin{equation}
\label{eq:ph2} \Gamma_{\beta 2}  = \frac{2 \pi}{\hbar} \sum_{k_1,
k_2,k_3,k_4} |\langle L_{k_1 \uparrow} R^\dagger_{k_2 \uparrow}
L_{k_3 \downarrow} R^\dagger_{k_4 \downarrow} H_\beta\rangle|^2
\delta(\xi_{k_1}+\xi_{k_3}-\xi_{k_2}-\xi_{k_4})= \frac{2 \pi }{
\hbar} \frac{\beta^2}{16 \pi^2 T_K^2} \int_{-eV}^{eV} d \epsilon
(V-\epsilon)(V+\epsilon).
\end{equation}\endwidetext
\noindent Here we used $\sum_k = \nu \int d\xi_k$, $\langle L_{k
\sigma} L^\dagger_{k' \sigma'} \rangle= \delta_{k k'}\delta_{\sigma
\sigma'} [1-f_d(\xi_k)]$, $\langle R^\dagger_{k \sigma} R_{k'
\sigma'} \rangle= \delta_{k k'}\delta_{\sigma \sigma'} f_s(\xi_k)$,
and set $T=0$. $\epsilon $ denotes the energy transfer from the
spin-up to the spin-down particle,
and the factors $V-\epsilon$ and $V+\epsilon$ originate from the integrations
over the initial energies of the spin-up and spin-down R-movers, respectively.
Performing the $\epsilon$-integration, we obtain
$I_{\beta 2 }= \frac{e^2}{h} \frac{2}{3} \bigl( \frac{V}{T_K}
\bigr)^2 V \beta^2$.

In a similar manner, one finds that inelastic backscattering processes
of a single R-mover
give a contribution to $I_b$ which is of the form $I_{\beta 1} =4
\Gamma_{\beta 1} e = \frac{e^2}{h} \frac{1}{6} \bigl(
\frac{V}{T_K} \bigr)^2 V \beta^2$.
The factor of $4$ reflects spin as well as the fact that the particle-hole pair can be
created either in the source or in the drain [see Fig.~\ref{fg:1}(c)].
$\Gamma_{\beta 1}$ is obtained by replacing $R^\dagger_{k_4 \downarrow} \rightarrow
L^\dagger_{k_4 \downarrow}$ in Eq.~(\ref{eq:ph2}) for
$\Gamma_{\beta 2}$. Note that the ratio of the phase-space factors
for inelastic backscattering of two movers vs.\ a single mover is
$\Gamma_{\beta 2}/\Gamma_{\beta 1}=8$.

The backscattering current due to elastic processes [see
Fig.~\ref{fg:1}(b)] follows from the elastic term in the
Hamiltonian Eq.~(\ref{eq:hfp}), which takes the form $H_\alpha =
-\frac{\alpha}{4 \pi \nu T_K}\sum_{k,k' \sigma}^{a,b=L,R}
(\xi_k+\xi_{k'}) a^\dagger_{k \sigma} b_{k' \sigma}$ in terms of
$L_{k\sigma}$ and $R_{k\sigma}$. This contains processes in which
at most one mover is backscattered. The corresponding elastic
contribution to $I_b$ is given by $I_\alpha =2 \Gamma_\alpha e =
\frac{e^2}{ h }\frac{1}{6} \bigl( \frac{V}{T_K} \bigr)^2 V
\alpha^2$, where the factor of two originates from spin.

Since the scattering events have rates $\propto \bigl( \frac{V}{T_K}
\bigr)^2 \ll 1$ and are thus rare, they are uncorrelated. For this
reason, the total shot noise $S =2e (I_\alpha + I_{\beta 1}+2
I_{\beta 2} )$ contains independent contributions from each process.
Using Eq.~(\ref{eq:qef}) and $ \alpha= \beta$, we recover the
effective charge
\begin{equation}
\label{eq:den} \frac{e^*}{e} = \frac{\frac{
\alpha^2}{6}+\frac{\beta^2}{6}+2\frac{2 \beta^2}{3}
}{\frac{\alpha^2}{6} +\frac{\beta^2}{6}+\frac{2 \beta^2}{3}
}=\frac{5}{3}.
\end{equation}

So far, we have derived this universal value of $e^*$ for systems
which reach the maximal unitary limit as $T\to0$. A necessary
condition for this to happen is that the system respects the
following symmetries:
(i)~$SU(2)$ spin symmetry, requiring zero magnetic field $\delta_h =
{g \mu_B H}/{T_K}=0$;
(ii)~particle-hole symmetry leading to the absence of potential
scattering, $\delta_r=0$;
 (iii) LR  symmetry, requiring dot-lead tunneling $t_{L,R}$ and capacitive couplings which
are equal for left and right lead. Deviations from $t_L=t_R$ imply
$\delta_\theta\neq 0$, where $\delta_\theta = 2 \theta - \pi/2$
with $\theta \equiv \arctan|t_R/t_L|$, $0 \leq \theta \leq \pi/2$.
Asymmetric capacitances may shift the position of the resonance
level further. We quantify this shift by a parameter $\gamma$
satisfying $E_F = \frac{\mu_s+\mu_d}2 + V \gamma$.

Some theoretical approaches artificially break these symmetries in
order to arrive at solvable models. E.g., the Schiller-Hershfield
version of the Toulouse solution~\cite{Schiller98} breaks $SU(2)$
spin symmetry as well as the LR-symmetry and indeed, we find that
it would predict $e^*=2e$. Slave-boson mean field theory neglects
two-particle scattering and breaks particle-hole
symmetry~\cite{Lopez05}. Since it leads to a self-consistent
single-electron description in terms of a resonance-level model,
one necessarily has $e^*/e=1$.

\begin{figure}[ht]
\begin{center}
\includegraphics*[width=60mm]{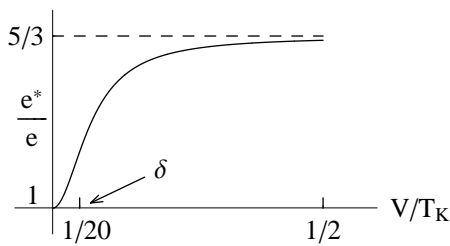}
\caption{Effective charge $e^*$ vs.\ voltage for broken LR tunneling symmetry
$\delta = \delta_\theta
=1/20$. The crossover from $e^*/e=1$ to $e^*/e \approx 5/3 $ occurs
when $V/T_K$ exceeds $\delta$. \label{fg:2}} {\vspace{-0.5cm}}
\end{center}
\end{figure}

{\it Realistic quantum dots}.---The maximal unitary limit is also
not easily accessible in experiment due to residual
symmetry-breaking perturbations~\cite{Wielshort00}. Such
perturbations lead to a backscattering current linear in $V$,
which dominates at low voltages and implies $e^*=e$. However, the
previously discussed ($\alpha$ and $\beta$) processes grow as
$V^3$ and will thus dominate at sufficiently high voltages,
leading to a crossover of $e^*$ to a value close to the universal
value $\frac{5}{3}e$. To quantify this scenario, we note that the
backscattering current $\propto V$, $I_b = 2 \frac{e^2}{h} V[1 -
\sin^2(2 \theta) \frac{1}{2} \sum_{\sigma = \pm } \sin^2
\delta_\sigma^{\rm{el}}]$~\cite{Glazman05}, is determined by the
LR-asymmetry $\delta_\theta$ and by the electronic phase shift
$\delta_\sigma^{\rm{el}} = \pi/2-\delta_r- \sigma  \delta_h$. The
latter differs by $\pi/2$ from the $\psi$-particles phase shift,
$-\delta_r- \sigma \delta_h$. This phase shift can be included by
adding to the Hamiltonian a local term, $H_{\rm{loc}} = \sum_{k k'
\sigma} \frac{\delta_r + \delta_h \sigma}{\pi \nu} \psi_{k
\sigma}^\dagger \psi_{k' \sigma}$. (A global magnetic field has a
similar contribution to the phase shift through a Hartree
treatment of the interaction.) If the dot is close to unitarity,
$\protect{\delta_\theta, \delta_h ,\delta_r \ll 1}$, we have the
expansion $I_b =2 \frac{e^2}{h} V \delta^2$, where $\delta^2 =
\delta_\theta^2 + \delta_h^2 + \delta_r^2$. Thus, this
contribution to backscattering becomes negligible once
\begin{equation}
\label{eq:con}
V^* = T_K \max \{ \delta_\theta, \delta_h, \delta_r
\} \ll V \ll T_K,
\end{equation}
and the detailed crossover of $e^*$ takes the form
\begin{equation}
\label{eq:cor} \frac{e^*}{e}=\frac{ \delta^2 V+\frac{5}{3}({V^3}/{2
T_K^2})}{
\delta^2 V+({V^3}/{2 T_K^2})},
\end{equation}
(with $\alpha = \beta  = 1$), as shown in Fig.~\ref{fg:2}.

It may be useful to note for experimental tests of our predictions
that for voltages $V\gg V^*$, it should be possible to subtract
explicitly the {\it identical} contributions $\propto V$ in the
noise $S/2e$ and in the backscattering current, cf.\ Eq.\ (\ref
{eq:exp53}) below. In this way, one isolates the terms $\propto
V^3$ and recovers the universal value $5/3$ even in the presence
of symmetry-breaking perturbations.

We remark that strictly speaking, symmetry-breaking perturbations
also affect the terms $\propto V^3$. These corrections appear at
yet higher order [e.g. ${\cal O}(\beta^2 \delta_r^2)$]. Evaluating
these corrections within the Keldysh approach for weak symmetry
breaking, we obtain
\begin{equation}
\label{eq:exp53} \frac{S/2e -\delta^2 g_0 V}{I_b-  \delta^2 g_0
V}-\frac{5}{3} \approx \sum_{i=r,h,\theta} c_i \delta_i^2
+c_\gamma \gamma^2 + \mathcal{O}(\delta^3),
\end{equation}
where $c_r, c_h, c_\theta$ and $c_\gamma$ are numbers of ${\cal O}(1)$ \cite{Remark06b}.

{\it Summary}.---We discussed the effective backscattering charge
measured by shot noise in Kondo quantum dots near perfect
transmission. The result $e^* = \frac{5}{3} e$ reflects the fact
that transport is mediated by a combination of one and two
particle scattering processes. We argued that even for real
quantum dots where most of the symmetries which are often assumed
in theoretical models are broken, the universal behavior can be
seen at voltages larger than a voltage scale $V^*$ reflecting the
strength of symmetry breaking, but smaller than $T_K$.

We acknowledge useful discussions with  A.~M.~Finkel'stein, L.\
Glazman, M.\ Heiblum, and O.\ Zarchin. Special thanks to A.\ Golub
and D.\ Meidan. This research was supported by DIP (YO and FvO),
ISF, BSF (YO), Sfb 658 (FvO), and the Studienstiftung d.\ dt.\
Volkes (JK).


\begin{thebibliography}{15}
\bibitem[{Rem({\natexlab{c}})}]{Remark06b}
\bibinfo{note}{Assuming that the fixed point Hamiltonian of Eq.~(\ref{eq:hfp})
  is defined for band width $D \ll T_K$ and neglecting terms of ${\cal O}
  (D/T_K)$ we find $c_\theta=c_\gamma = -4/3, \; c_h=c_r=223/144$}.
\end{thebibliography}
\end{document}